\documentclass[a4paper]{article}

\usepackage[english]{babel}
\usepackage[utf8]{inputenc}
\usepackage{amsmath}
\usepackage{graphicx}
\usepackage[colorinlistoftodos]{todonotes}
\usepackage{csquotes}
\usepackage{multicol}
\usepackage{scrextend}
\usepackage{endnotes}

\usepackage{url}

\title{Critique of J. Kim's ``\textbf{P} is not equal to \textbf{NP} by Modus Tollens''}

\author{Dan Hassin, Adam Scrivener, and Yibo Zhou\vspace{8pt} \\Department of Computer Science\\University of Rochester\\Rochester, NY 14627, USA}

\date{\today}

\begin{document}

\maketitle

\begin{abstract}
This paper is a critique of version three of Joonmo Kim's paper entitled ``$\textbf{P} \ne \textbf{NP}$ by Modus Tollens.'' After summarizing Kim's proof, we note that the logic that Kim uses is inconsistent, which provides evidence that the proof is invalid. To show this, we will consider two reasonable interpretations of Kim's definitions, and show that ``$\textbf{P} \ne \textbf{NP}$'' does not seem to follow in an obvious way using any of them. 
\end{abstract}

\section{Introduction}
The abstract of Kim's paper \cite{Kim2014} is as follows:

\blockquote{
An artificially designed Turing Machine algorithm $\textbf{M}^o$ generates the instances of the satisfiability problem, and check their satisfiability. Under the assumption $\textbf{P} = \textbf{NP}$, we show that $	\textbf{M}^o$ has a certain property, which, without the assumption, $	\textbf{M}^o$ does not have. This leads to $\textbf{P} \ne \textbf{NP}$ by modus tollens. }

In this paper we will critique Kim's proof that $\textbf{P} \ne \textbf{NP}$, a proof that claims to solve the famous $\textbf{P} $ vs. $ \textbf{NP}$ problem, widely agreed to be the most important unsolved problem in computer science. We will begin by outlining Kim's paper in detail. Then we will examine several problems with his paper, including the many ambiguous definitions, wordings, and explanations throughout his paper. We will address both of the possible interpretations of his most ambiguous definition, and describe how either interpretation arrives at a contradiction. Finally, we provide a comment on the invalidity of the final commentary of Kim's paper.

\section{Kim's argument}
Most of the arguments in Kim's paper are not rigorous, some constructions are not shown, his definitions are not precise, and we believe his paper contains many notational mistakes. For this reason, this section will restate his paper in detail as we understand it so that it may be refuted rigorously.

\subsection{Cook-Levin theorem}

By the Cook-Levin theorem, we can construct a series of Boolean clauses based on the action of any arbitrary Turing machine on a given input $x$, such that the series of clauses is satisfiable if and only if it accepts $x$. Together, Kim calls this unit of clauses \textbf{c}. In the proof of this theorem given by Garey and Johnson, these Boolean clauses in \textbf{c} are grouped into $G_1$, $G_2$, $G_3$, $G_4$, $G_5$, and $G_6$ by the different parts of the computation that they enforce. One group of particular interest to Kim is $G_4$, which enforces that ``at time 0, the computation is in the initial configuration of its checking stage for input $x$'' \cite{Garey1979}. The rest of the groups of clauses, $G_1$, $G_2$, $G_3$, $G_5$, and $G_6$, are each concerned with the run of the machine, its transitions, and that it ends at a final state \cite{Garey1979}.

For a given \textbf{c}, he lets \textbf{c}$^x$, what he calls the ``input-part,'' be $G_4$, the group that is concerned with asserting the initial configuration. He lets \textbf{c}$^r$, the ``run-part,'' be the rest of the clauses, those concerned with the run of the machine.

\subsection{Construction of \textbf{M}}

$\textbf{Definition:}$ An \emph{accepting computation} of a Turing machine $T$ on input $y$ is a finite sequence of configurations and transitions of $T$ (starting at the initial state of $T$ and with $y$ on its tape) that ends in an accepting state of $T$.\\

\noindent Kim proposes a Turing machine algorithm \textbf{M}, which has in its code a finite list $\textbf{c}_1, \textbf{c}_2, ..., \textbf{c}_n$ of Boolean formulas, each which have been encoded from arbitrary accepting computations from arbitrary Turing machines, using the Cook-Levin construction described above. For each $\textbf{c}_j$ in this list, we can strip its ``input-part'' to obtain the list of ``run-parts'' \textbf{c}$^r_1$, \textbf{c}$^r_2$, ..., \textbf{c}$^r_n$. Call this list of ``run-parts'' from arbitrary machines $C^r_\textbf{M}$.

Kim notes that we can construct countably many \textbf{M}'s, $\textbf{M}_1, \textbf{M}_2, \textbf{M}_3, ..., \textbf{M}_i$, where each $\textbf{M}_i$ includes in its program code the list of Boolean formulas $C^r_{\textbf{M}_i}$, as described above. Kim defines the behavior of a given $\textbf{M}_i$ as follows.

\vspace{2mm}
\begin{addmargin}[30pt]{0pt}
$\textbf{M}_i = $ ``On input $y$:

\begin{addmargin}[7pt]{0pt}

\begin{enumerate}
\item \vspace{1mm} Using the Cook-Levin construction, compute the Boolean clause \textbf{c}$^y$ (i.e., $G_4$, the clause that enforces the initial machine
state of $\textbf{M}_i$ and the placement of input $y$ on its tape.)\endnote{The Garey and Johnson construction that Kim uses for the \textit{SAT} reduction, even for group $G_4$, is reliant on a machine so that its initial state can be encoded: ``at time 0, the computation is in the initial configuration of its checking stage for input $x$'' \cite{Garey1979}.

However, Kim does not specify which machine to use in this construction of $\textbf{c}^y$. So that the machine code for $\textbf{M}_i$ is well defined, we use $\textbf{M}_i$ itself as the machine used in the $\textbf{c}^y$ construction, but the following argument will show that in fact the choice of machine does not matter. (And since Garey and Johnson's construction is not the only conceivable construction that maps machine actions to clauses (with separate ``input'' and ``run'' parts) as per the Cook-Levin theorem, our proof is independent of the details of that specific construction.)

Suppose that a given Boolean formula $\textbf{c}^y$ was constructed from machine $M$ on input $y$, and that an arbitrary Boolean formula $\textbf{c}^r$ was constructed from machine $M^r$.

\begin{itemize}
	\item If the only relevant information in $\textbf{c}^y$, even being constructed from $M$ specifically, is the placement of $y$ on the tape (i.e., that the construction encodes all initial states for all machines the same way), then indeed concatenating $\textbf{c}^r$ from $M^r$ will produce the same $\textbf{c}$ that the construction would produce for $M^r$ on input $y$.

	\item If the information in $\textbf{c}^y$ uses specific elements of $M$, like $M$'s start state, it is still possible that the clauses of $\textbf{c}^r$ are ``compatible'' with $\textbf{c}^y$ (i.e., the construction produces the same $\textbf{c}^y$'s for both $M$ on input $y$ and $M^r$ on input $y$), and so it is \emph{possible} that $(\textbf{c}^y \land \textbf{c}^r)$ is satisfiable. 

	\item If the information in $\textbf{c}^y$ is absolutely specific to $M$ and is entirely ``incompatible'' with any other machine Boolean encodings, then no \textbf{c} will be satisfiable, meaning that no $\widehat{ac}_{\textbf{c}}$ and thus no $\textbf{c}^o$ (explained in Section 2.3) can exist. This would make the antecedent in $(P_2 \Rightarrow P_3)$ in Section 2.4 false, which would make the statement true, and thus the logical model is valid regardless of the truth value of $P_1$ ($\textbf{P} = \textbf{NP}$.)
\end{itemize}
}

\item \vspace{-2mm} For each \textbf{c}$_j^r \in C^r_{\textbf{M}_i}$:
\begin{enumerate}
\item \vspace{-2mm} Concatenate \textbf{c}$_j^r$ with \textbf{c}$^y$ to form \textbf{c}$_j$.
\item \vspace{-1mm} Give \textbf{c}$_j$ to a \emph{SAT}-solver module, and increment a counter if it returns that \textbf{c}$_j$ is satisfiable (i.e., if it is a valid accepting computation with input $y$.)
\end{enumerate}
\item \vspace{-2mm} Accept if the number accumulated by the counter is odd.''
\end{enumerate}
\end{addmargin}

\end{addmargin}

\vspace{10pt}

\noindent
Let $C_{\textbf{M}_i , y}$ be the list of each \textbf{c} that appears in the run of $\textbf{M}_i$ on input $y$. More formally, $C_{\textbf{M}_i , y} = \textbf{c}_1, \textbf{c}_2, ..., \textbf{c}_n$ such that $\textbf{c}_j = (\textbf{c}^y \land \textbf{c}^r_j)$, where $\textbf{c}^r_j \in C^r_{\textbf{M}_i}$. \\

\subsection{Defining $\textbf{M}^o$}

Kim introduces the idea of a \emph{particular transition table} of a Turing machine, which he says is a transition table that ``may produce an accepting computation by running on a Turing Machine'' \cite{Kim2014}. He observes that ``each of all accepting computations may have its particular transition table, i.e., the table can be built by collecting all the distinguished transitions from the computation, where we know that a computation is a sequence of the transitions of configurations of a Turing Machine'' \cite{Kim2014}.

Kim then proposes the machine $\textbf{M}^o \in \{\textbf{M}_1, \textbf{M}_2, ..., \textbf{M}_i\}$ such that, for some input $y$, there exists a \textbf{c}$^o \in C_{\textbf{M}^o , y}$ that describes an accepting computation on input $y$ (call this accepting computation $\widehat{ac}_{\textbf{c}^{o}}$) for which there exists a particular transition table \textbf{t} which is also a particular transition table for the accepting computation of the run of $\textbf{M}^o$ on $y$ (call this $\widehat{ac}_{\textbf{M}^{o}}$.) Note, rather importantly, that $\widehat{ac}_{\textbf{c}^{o}}$ and $\widehat{ac}_{\textbf{M}^{o}}$ are both accepting computations with respect to input $y$. 

\subsection{``$\textbf{P} \ne \textbf{NP}$''}

\textbf{Definition:} \cite{Kim2014} A particular transition table \textbf{t} is $D_{sat}$ if it ``deterministically describes $\textbf{M}^{o}$'s transitions and the SAT-solver module in $\textbf{M}^{o}$ runs deterministically in a poly-time for the length of \textbf{c}.'' \\

\noindent Kim's proof of $\textbf{P} \ne \textbf{NP}$ is as follows.

\blockquote{
$P_1$: $\textbf{P} = \textbf{NP}$,\\
$P_2$: $\textbf{M}^o$ exists,\\
$P_3$: there exists \textbf{t}, which is $D_{sat}$.

By modus tollens, $(P_1 \Rightarrow (P_2 \Rightarrow P_3)) \land (\neg (P_2 \Rightarrow P_3))$ may conclude $\neg P_1$. \cite{Kim2014}
}

He argues that $P_1 \Rightarrow (P_2 \Rightarrow P_3)$ because, if such an $\textbf{M}^o$ exists, by definition a \textbf{t} must exist, and that \textbf{t} is $D_{sat}$, because if $\textbf{P}=\textbf{NP}$, the $SAT$-solver module (which is known to be \textbf{NP}-complete) would run in deterministic polynomial time.

This argument is sufficient to show that $P_1 \Rightarrow (P_2 \Rightarrow P_3)$. All Kim must show now is that $P_2 \Rightarrow P_3$ results in a contradiction, thus showing $\neg (P_2 \Rightarrow P_3)$, and proving by modus tollens $\neg P_1$. But first, he argues (unnecessarily) that $P_2$ is true. And in fact, he is actually arguing that $P_2 \land P_3$ is true. The following argument \cite{Kim2014}, while irrelevant, is shown for completeness.

\blockquote{
We can show that $P_2$ is true, as follows. For any chosen $\textbf{c}^o$, build two non-deterministic particular transition tables for $\widehat{ac}_{\textbf{M}^o}$ and $\widehat{ac}_{\textbf{c}^o}$ separately, and then merge the two so that one of the two computations can be chosen selectively from the starting state during the run. $\textbf{M}^o$ may exist by this \textbf{t}, which is $ND_{sat}$.
}

Kim provides very little to explain precisely what this ``merging process'' of tables is. Below is a construction of what we assume his merging process to be. Consider transition tables $\delta_M$ and $\delta_{M'}$ for machines $M$ and $M'$ respectively: {\small (For brevity, head movements and writes to the tape are omitted; only state transitions are shown.)}

\begin{multicols}{2}

\begin{center}
\begin{tabular}{l | *{4}{c}r}
$\delta_{M}$ &  {\tt a} & {\tt b} & $\cdots$ \\
\hline
$q_0$     &  $q_w$  &   $q_x$  \\
$q_1$     &  $q_y$  &   $q_z$   \\
$\vdots$      &    & &   $\ddots$   \\
\end{tabular}

\begin{tabular}{l | *{4}{c}r}
$\delta_{M'}$ &  {\tt a$'$} & {\tt b$'$} & $\cdots$ \\
\hline
$q_0'$     &  $q_w'$  &   $q_x'$  \\
$q_1'$     &  $q_y'$  &   $q_z'$   \\
$\vdots$  &           &  & $\ddots$ \\ 
\end{tabular}
\end{center}

\end{multicols}
\noindent Using $\delta_M$ and $\delta_{M'}$, we can produce the non-deterministic particular transition table $\delta_{M,M'}$ that ``describes,'' or that can ``produce'' both $\widehat{ac}_{M}$ and $\widehat{ac}_{M'}$, as generated by $\delta_{M,M'}$, for arbitrary input $y$:

\begin{center}
\begin{tabular}{l | *{8}{c}}
$\delta_{M,M'}$ & $\epsilon$ & {\tt a} & {\tt b} & $\cdots$ & {\tt a$'$} & {\tt b$'$} & $\cdots$ \\
\hline
$q_{\text start}$  &  $\{q_0, q_0'\}$ & $\emptyset$ & $\emptyset$ & & $\emptyset$ & $\emptyset$  \\
$q_0$   & $\emptyset$ &  $\{q_w\}$  &   $\{q_x\}$ & & $\emptyset$ & $\emptyset$ \\
$q_1$   & $\emptyset$  &  $\{q_y\}$  &   $\{q_z\}$ & & $\emptyset$ & $\emptyset$   \\
$\vdots$  & & & & $\ddots$ & \\
$q_0'$  & $\emptyset$ & $\emptyset$ & $\emptyset$  & &  $\{q_w'\}$  &   $\{q_x'\}$  \\
$q_1'$  & $\emptyset$ & $\emptyset$ & $\emptyset$  & & $\{q_y'\}$  &   $\{q_z'\}$   \\
$\vdots$  & & & & & & & $\ddots$ \\
\end{tabular}
\end{center}

\subsection{Contradiction argument}

Kim's proof by contradiction to prove $\neg (P_2 \Rightarrow P_3)$ is as follows. By way of contradiction, he assumes $(P_2 \Rightarrow P_3)$ to be true, i.e., that ``if $\textbf{M}^{o}$ exists then there exists \textbf{t}, which is a $D_{sat}$ particular transition table for both $\widehat{ac}_{\textbf{M}^o}$ and $\widehat{ac}_{\textbf{c}^o}$'' \cite{Kim2014}.  He then claims that, since the same transition table \textbf{t} can generate both $\widehat{ac}_{\textbf{M}^o}$ and $\widehat{ac}_{\textbf{c}^o}$, which share the same input $y$, ``it is concluded that both $\widehat{ac}_{\textbf{M}^o}$ and $\widehat{ac}_{\textbf{c}^o}$ are exactly the same computation, i.e., all the transitions of the configurations of $\widehat{ac}_{\textbf{M}^o}$ and those of $\widehat{ac}_{\textbf{c}^o}$ are exactly the same'' \cite{Kim2014}.

Now, he lets $i$ be the number of transitions between configurations in $\widehat{ac}_{\textbf{M}^o}$,  $j$ the number of clauses of $\textbf{c}^o$, and $k$ the number of transitions between configurations in $\widehat{ac}_{\textbf{c}^o}$.

He argues that during the run of $\textbf{M}^o$ on input $y$, all the clauses of $\textbf{c}^o$ will have to be loaded on the tape of $\textbf{M}^o$, as well as the clauses of all other \textbf{c}'s $\in C_{\textbf{M}^o, y}$, so $i > j$. And, since each transition of an accepting computation is described by more than one clause \cite{Garey1979}, we conclude $j > k$, and thus $i > j > k$.

However, Kim argues that a contradiction arises here. The previous conclusion that $\widehat{ac}_{\textbf{M}^o}$ and $\widehat{ac}_{\textbf{c}^o}$ are exactly the same computation would imply that $i = k$, which contradicts $i > j > k$. Thus, he claims $\neg (P_2 \Rightarrow P_3)$.

\section{Critique}
During our analysis of his argument, we identified several flaws in Kim's proof which we critique here in detail.

\subsection{Invalidity of logical argument}

Kim's argument centers around the definition of ${D_{sat}}$, as well as this fact: if $\textbf{P} = \textbf{NP}$ then the particular transition table that is implied by $\textbf{M}^{o}$'s existence is $D_{sat}$.
Kim then attempts to arrive at a contradiction by showing that such a particular transition table cannot exist. However, in his proof by contradiction, he does not use the fact that \textbf{t} is $D_{sat}$, so the assumption (that if there exists an $\textbf{M}^{o}$ then there exists a \textbf{t} that is $D_{sat}$) is equivalent to ($\textbf{M}^{o}$ exists) $\Rightarrow$ (\textbf{t} exists). Note that, by definition, $\textbf{M}^{o}$ exists if and only if \textbf{t} exists. Therefore, Kim cannot possibly prove that ($\textbf{M}^{o}$ exists) $\Rightarrow$ (\textbf{t} exists) is false. This fact provides evidence that his proof must be invalid, which we will presently show. 


\subsection{Ambiguities with accepting computations and particular transition tables}

An error arises in Kim's final contradiction that $P_{2}$ does not imply $P_{3}$, namely that since $\textbf{M}^{o}$ exists, a $\textit{D}_{sat}$ particular transition table of both $\widehat{ac}_{\textbf{c}^{o}}$ and $\widehat{ac}_{\textbf{M}^{o}}$ exists. Kim argues that the existence of this particular transition table implies that $\widehat{ac}_{\textbf{c}^{o}}$ and $\widehat{ac}_{\textbf{M}^{o}}$ are equivalent accepting computations.

Here, Kim's definition of an \emph{accepting computation} is of crucial importance. Michael Sipser \cite{Sipser2013} offers the following definition of an \emph{accepting computation history}:

\blockquote{
	Let $M$ be a Turing machine and $w$ an input string. An \emph{accepting computation history} for $M$ on $w$ is a sequence of configurations, $C_1, C_2, ..., C_l$, where $C_1$ is the start configuration of $M$ on $w$, $C_l$ is an accepting configuration of $M$, and each $C_i$ legally follows from $C_{i-1}$ \textbf{according to the rules of $\boldsymbol{M}$}. (Emphasis added.) 
}

\noindent Although Sipser refers to an accepting computation \textit{history}, we infer from Kim's own paper that this definition is equivalent to simply \textit{accepting computation}: ``...we know that a computation is a \textbf{sequence of the transitions of configurations} of a Turing Machine'' (emphasis added) \cite{Kim2014}.
 
Note that Sipser's definition suggests, as would common intuition, that an accepting computation relies on the transition table of the given machine running it. However, Kim is vague in describing how particular transition tables and accepting computations relate. One could interpret it in one of two ways. Either,
\begin{enumerate}
\item An accepting computation is produced by a given Turing machine and its own transition table.
\item An accepting computation can be produced by a given particular transition table, not necessarily that of the original machine, that can describe each transition between configurations.
\end{enumerate}

\noindent We believe that an error arises when Kim operates under the first interpretation for his claim that $i > j > k$, and the second for the $i=k$ claim. To produce a consistent and coherent proof, the paper can only operate under one interpretation. In the following sections, we will address both interpretations independently and show that under either one, his contradiction is invalid.

\subsubsection{First interpretation}
By this interpretation, $\widehat{ac}_{\textbf{M}^{o}}$ and $\widehat{ac}_{\textbf{c}^{o}}$ are accepting computations from different Turing machines entirely, which behave in very different ways. ${\textbf{M}^{o}}$, on input $y$, concatenates its own $\textbf{c}^y$ with each of the Boolean formulas in $C^r_{\textbf{M}^o}$, then runs a \textit{SAT}-solver module on each $\textbf{c}$, counting the $\textbf{c}$'s that are accepted. On the other hand, each $\widehat{ac}_{\textbf{c}}$ is just an arbitrary accepting computation of some Turing machine $\textbf{M}$ on an input $y$. Under this interpretation, it is not obvious that any $\widehat{ac}_{\textbf{M}^{o}}$ is equivalent to any $\widehat{ac}_{\textbf{c}^{o}}$. The argument that Kim gives as proof that some $\widehat{ac}_{\textbf{M}^{o}}$ is equivalent to some $\widehat{ac}_{\textbf{c}^{o}}$ is that one can create a particular transition table that is a transition table for both $\widehat{ac}_{\textbf{M}^{o}}$ and $\widehat{ac}_{\textbf{c}^{o}}$. However, the ``merging'' technique that Kim uses to show that any \textbf{t} can be made from two transition tables can be shown to be invalid.

This technique produces a new transition table, which contains new states (as it must include the set of states from both machines) and possibly new alphabet characters. Thus, it cannot be said that the new particular transition table is the same transition function as either original machine, or even a ``compatible'' one, since it operates on a set of states that is different from the machine's original set of states, and would thereby be malformed.

Therefore, under this interpretation, Kim's argument that $i = k$ follows from there existing some \textbf{t} which can produce $\widehat{ac}_{\textbf{M}^{o}}$ and $\widehat{ac}_{\textbf{c}^{o}}$ is invalid, since $\widehat{ac}_{\textbf{M}^{o}}$ and $\widehat{ac}_{\textbf{c}^{o}}$ are computations produced by transition tables necessarily different from their original machines, and thus $i > j > k$ is correct, and there is no inconsistency.

\subsubsection{Second interpretation}
In the second interpretation, we will assume that accepting computations can be produced by the particular transition table $\textbf{t}$. Then we may conclude that since $\textbf{t}$ is a particular transition table for $\widehat{ac}_{\textbf{M}^{o}}$ and $\widehat{ac}_{\textbf{c}^{o}}$, then $\widehat{ac}_{\textbf{M}^{o}}$ and $\widehat{ac}_{\textbf{c}^{o}}$ are equivalent \textbf{as accepting computations produced by a particular transition table for some input $\boldsymbol{y}$}. Note that these accepting computations are not necessarily the same as the accepting computations produced by their respective Turing machines' transition tables. So, when Kim concludes that the number of transitions in $\widehat{ac}_{\textbf{M}^{o}}$ must be larger than the number of transitions in $\widehat{ac}_{\textbf{c}^{o}}$ as a contradiction, he is no longer comparing the same accepting computations, so that fact is not contradictory. Given the nature of $\textbf{M}^o$ and $\textbf{c}^o$, $i > j > k$ does not follow, because $\widehat{ac}_{\textbf{M}^{o}}$ and $\widehat{ac}_{\textbf{c}^{o}}$ are indeed the same computations produced by \textbf{t}, and analysis based on their original respective machines does not apply.

\subsection{Comment on Kim's ``commentary''}

At the end of his paper, Kim verifies that his given proof could not also be used to prove that $\textbf{P} = \textbf{NP}$. This verification is very brief and relies heavily on the assumption that the proof that his paper presents makes accurate assumptions and logical inferences. It is essentially a retelling of his argument with reversed assumptions and conclusions. Clearly, assuming that his original proof is correct, it can be used to refute the possibility of proving the opposite statement, but it does not tell us anything about the validity of the original proof. 

\section{Conclusion}
From our interpretation of Kim's paper, the main problems stem from a severe lack of rigor, numerous misunderstandings, and occasional inconsistencies in his definitions. In his main argument, he derives a contradiction from the properties of a Turing machine and the $D_{sat}$ property of a particular transition table. In our main argument, we point out that there is an inconsistency here that renders the main proposition of his supposed contradiction invalid in the two possible interpretations of his definition regarding \textit{accepting computation}. 

\subsection*{Acknowledgments}

We thank Lane A. Hemaspaandra and Joe Izraelevitz for helpful comments on a preliminary draft of this critique. All claims, opinions, and errors in the present, substantially revised critique are the sole responsibility of the authors.


\theendnotes


\bibliography{bib}{}
\bibliographystyle{plain}

\end{document}